\documentclass[12pt]{iopart}
 \input{epsf.sty}
 \input{epsf.tex}
 \eqnobysec

\newcommand{\beq}{\begin{equation}}
\newcommand{\eeq}{\end{equation}}
\newcommand{\bea}{\begin{eqnarray}}
\newcommand{\eea}{\end{eqnarray}}

\def\sqr#1#2{{\vcenter{\hrule height.#2pt \hbox{\vrule width.#2pt
       height#1pt\kern#1pt \vrule width.#2pt} \hrule height.#2pt}}}

\def\laq{\raise 0.4ex\hbox{$<$}\kern -0.8em\lower 0.62
ex\hbox{$\sim$}}
\def\gaq{\raise 0.4ex\hbox{$>$}\kern -0.7em\lower 0.62
ex\hbox{$\sim$}}

\def \ra {\rightarrow}

\def \La {\Lambda}

\def \ap {\alpha^{\prime}}

\def \om {\omega}

\def \noi {\noindent}

\begin{document}


\begin{flushright}
BA-TH/99-367\\
hep-th/0004149
\end{flushright}

\vspace*{0.8truein}

{\Large\bf\centering\ignorespaces
String Cosmology\\ 
\bigskip
versus Standard and Inflationary Cosmology
\vskip2.5pt}
{\dimen0=-\prevdepth \advance\dimen0 by23pt
\nointerlineskip \rm\centering
\vrule height\dimen0 width0pt\relax\ignorespaces

\vspace{0.8 cm}
M. Gasperini
\par}
\vspace{0.5 cm}
{\small\it\centering\ignorespaces
Dipartimento di Fisica, Universit\`a di Bari,\\
Via G.  Amendola 173, 70126 Bari, Italy \\
\vspace{0.2 cm}
and\\
\vspace{0.2 cm}
Istituto Nazionale di Fisica Nucleare,  Sezione di Bari, Bari, Italy\\
\par}

\par
\bgroup
\leftskip=0.10753\textwidth \rightskip\leftskip
\dimen0=-\prevdepth \advance\dimen0 by17.5pt \nointerlineskip
\small\vrule width 0pt height\dimen0 \relax

\vspace*{0.6truein}

\centerline{\bf Abstract}

\noi
This paper presents a review of the basic, model-independent 
differences between the pre-big bang scenario, arising naturally in a  
string cosmology context,  and the standard inflationary scenario. 
We use an unconventional approach in which the introduction of 
technical details is avoided as much as possible, trying to focus the 
reader's attention on the main conceptual aspects of both scenarios. 
The aim of the paper is not to conclude in favour either of one or  
of the  other scenario, but to raise questions that are left to  
the reader's meditation. Warnings: the paper does not contain 
equations, and is not intended as a complete review of all aspects of 
string cosmology. 

\vspace{0.8cm}
\begin{center}
------------------------------  

\vspace{0.8cm}
To appear  in {\bf  Classical and Quantum Gravity}\\ 
(Topical Review Section)
\end{center}

\thispagestyle{plain}
\par\egroup

\vfill

\maketitle

\setcounter{page}{1}

\title[String cosmology]{String cosmology versus standard and
inflationary cosmology }

\author{M. Gasperini\dag\ddag
\footnote[3]{E-mail: gasperini@ba.infn.it.}}

\address{\dag Dipartimento di Fisica , Universit\`a di Bari, 
Via G. Amendola 173, 70126 Bari, Italy}

\address{\ddag Istituto Nazionale di Fisica Nucleare,  Sezione di Bari,
Bari, Italy}

\begin{abstract}
This paper presents a review of the basic, model-independent 
differences between the pre-big bang scenario, arising naturally in a  
string cosmology context,  and the standard inflationary scenario. We use 
an unconventional approach in which the introduction of technical details 
is avoided as much as possible, trying to focus the reader's attention on 
the main conceptual aspects of both scenarios. The aim of the paper is 
not to conclude in favour either of one or  of the  other scenario, but to 
raise questions that are left to  the reader's meditation. Warnings: the
paper does not contain equations, and is not intended as a
complete review of all aspects of string cosmology. 

\end{abstract}

~~~~~~~~~~~~~Preprint  BA-TH/99-367, 
~~~~~~~~~~~~~E-print Archives: hep-th/0004149


\section{Introduction}
\label{I}

The standard cosmological scenario \cite{Wein,0}, rightfully one of the best
celebrated conquests of the physics of the XX century, cannot be
extrapolated to arbitrarily high energy and curvature scales without
clashing with the singularity problem. 

A singularity, on the other hand,  often represents a signal that 
the physical laws  we are applying have been extrapolated outside
their domain of validity.  As a well known example, we may quote here the
case of the spectral energy distribution of radiation in thermal
equilibrium. By applying  the laws of classical physics one finds indeed 
the Rayleigh-Jeans spectrum, that diverges like $\om^3$ at high
frequency. By taking into account instead the appropriate quantum
corrections, this classical singularity is regularized by the bell-like
Planck distribution, $\om^4\left(e^{\om/T}-1\right)^{-1}$, as
illustrated in Fig. 1. 

\begin{figure}
\centerline{\epsfxsize=9.0cm
\epsffile{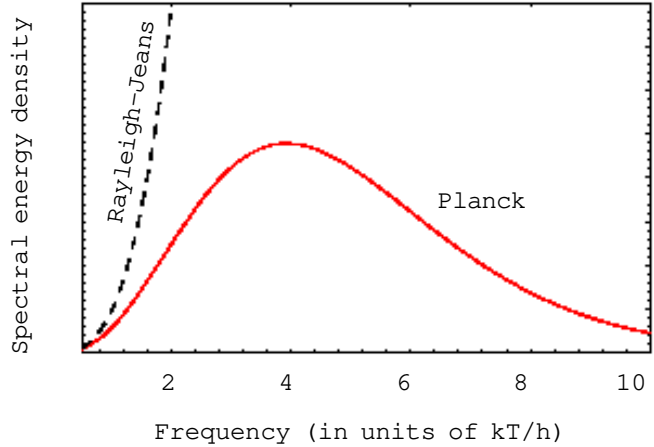}}
\caption{\sl The Planck distribution (full line) regularizes the classical
Rayleigh-Jeans prediction (dashed line), for the spectral energy density
of radiation in thermal equilibrium. The classical distribution is only 
valid at low enough frequency scales.}
\label{fig1}
\end{figure}

String theory suggests that the initial curvature singularity of the
standard cosmological scenario could be similarly regularized by a
bell-shaped curve, as qualitatively illustrated in Fig. 2. As we go back 
in time the curvature, instead of blowing up, could reach a maximum 
controlled by the string scale, and then decrease towards a state of 
very low curvature and weak coupling, approaching asymptotically the
so-called string perturbative vacuum. This behaviour, indirectly suggested
by  duality and thermodynamical  arguments \cite{1}, as well as,
independently, by  the motion of strings in rolling   backgrounds \cite{2}, is
naturally grounded on the duality symmetry of the cosmological string
effective action \cite{3},  and its possible implementation in the context of
a realistic model of the early Universe has a lot of dynamical and
phenomenological consequences \cite{4}.  

In the string comology scenario of  Fig. 2, the  big bang singularity
is replaced  by a phase of high but finite curvature. It comes thus  natural,
in  such a context, to call ``pre-big bang" the initial phase in which the 
curvature is growing, in contrast to the subsequent ``post-big bang" 
phase, with decreasing curvature and standard decelerated evolution. 

\begin{figure}[b]
\centerline{\epsfxsize=9.0cm
\epsffile{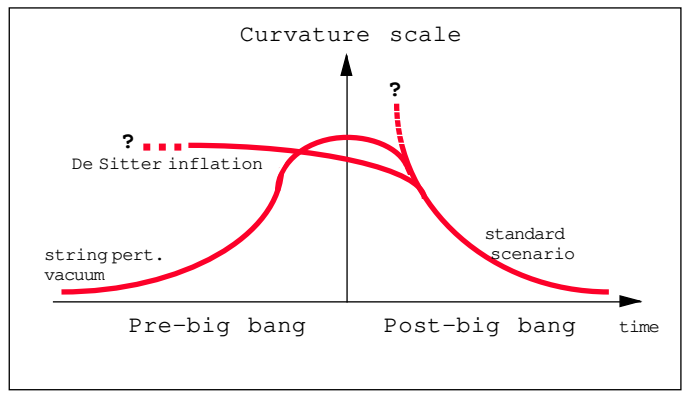}}
\caption{\sl Curvature scale versus time for standard, inflationary (de 
Sitter), and string cosmology models of the Universe. In the pre-big bang 
scenario  the classical curvature singularity is regularized, and the
standard  cosmological evolution is only valid at late enough time scales.}
\label{fig2}
\end{figure}

The conventional inflationary scenario \cite{0,4a}, on the other hand, 
suggests a picture of the early Universe which is approximately 
intermediate between the standard and the string cosmology one.  In that
context the  curvature, instead of blowing up, is
expected to approach a nearly  constant value, typical of a de Sitter-like
geometry (see Fig. 2). However, a de Sitter phase with 
exponential expansion at constant  curvature, implemented in the context
of the conventional, potential-dominated inflation,  cannot be extended
back in time for ever, as first discussed in  \cite{5}. Indeed, quoting Alan
Guth's recent survey of inflationary cosmology  \cite{6}:

\bigskip 

{\sl ``... Nevertheless, since inflation appears to be eternal 
only into the future, 
but not to the past, an important question remains open. How did 
all start? Although eternal inflation pushes this question far into the 
past, and well beyond the range of observational tests, the question 
does not disappear."}

\bigskip 

A possible answer to the above question, suggested by string cosmology 
and represented graphically in Fig. 2, is that all started from a state
approaching,  asymptotically, the flat, cold and  empty 
string perturbative vacuum. Even if this starting point is infinitely 
far into the past, however,   the initial state of our
Universe  might be non completely beyond the range of present
observations,  in contrast to the sentence quoted above, because the
starting point may affect the dynamics of the subsequent  inflationary
evolution. The initial curvature scale, for instance, is  constant or
decreasing according to the conventional inflationary  scenario,
while it is growing according to the pre-big bang scenario, and this  may
lead to important phenomenological consequences. 

In spite of the existence of a few particular examples \cite{6a}, an
unambiguous regularization of the curvature singularity (to all orders in
the string effective action), toghether with a complete description of the
transition from the pre- to the post-big bang regime, is not an easy
achievement (see however \cite{6b} for recent encouraging results). 
Also, it is fair to say that the pre-big bang models, at present, are not
free from other (more or less important) difficulties, that some aspects of
the pre-big bang scenario are still unclear, and that further work is
certainly needed for a final answer to all the difficulties. 

Assuming  that all the problems can (and will) be solved in a
satisfactory way, string cosmology will provide eventually a model of the
early Universe somewhat different, however, from the standard
inflationary picture.  The aim of this paper, therefore, is to  present in a
compact form a comparison (and a short discussion) of the way in which a
phase of inflation could be implemented within a cosmology based on the
string effective action, with respect to the standard cosmology based on
the Einstein equations. Basically all the differences arise, as we will see,
from the  fact that in string cosmology the Universe starts evolving from
an  initial state at very low curvature and weak coupling, while in 
conventional inflation the initial state is assumed to approach the 
Planckian, quantum gravity regime. 

The paper is organized as follows. We will discuss kinematical and 
dynamical differences in Sect. \ref{II}, 
quantum cosmology  differences in Sect. \ref{III}, 
and phenomenological  differences in Sect. \ref{IV}. 
Sect. \ref{V} is devoted to some concluding remarks.   We will avoid as
much as possible the  introduction of technical details -- all contained in
the papers quoted in the bibliography -- trying to emphasize the ideas and
the main physical aspects of the different inflationary scenarios. 

As already stressed in the Abstract, it seems appropriate to recall that
this paper is {\em not} intended as a complete review of all aspects of
string cosmology. Rather, the paper is narrowly focused on those aspects
where there is an important overlapping of methods and objectives and,
simultaneously, a strong contrast of basic assumptions, for the string
cosmology and the standard scenario. For a more exhaustive and
systematic approach, the interested reader is referred to the excellent
review paper devoted to a presentation and a technical discussion of all
presently existing (super)string cosmology models \cite{10a}, as well as
to two recent introductory lectures on the pre-big bang scenario
\cite{lec}. 

\section {Kinematical and dynamical differences}
\label{II}

The idea of inflation hystorically was born \cite{7} to solve the 
problem of monopoles that could be largely produced in the early 
Universe, at the energy scale  of grand unified theories ($GUT$). More
generally, inflation is now  understood as a period of accelerated
evolution that can explain why the  present Universe is so flat and smooth
over a so large scale of  distances \cite{0,4a}. 

For hystorical reasons, i.e. because inflation was first implemented as  
a period of supercooling of the Universe trapped in a ``false" vacuum 
state, in the context of $GUT$ phase transitions, inflation was first 
associated  \cite{7} to a phase of exponential 
expansion (in cosmic time) 
of the scale factor $a(t)$, corresponding to a de Sitter (or 
``quasi-de Sitter") geometrical state. But it was soon realized that any 
type of accelerated expansion ($\dot a >0$, $\ddot a >0$, where the dots 
denote differentiation with respect to cosmic time), can in principle 
solve the kinematic problems of the standard scenario \cite{8}. 

Actually, even accelerated contraction ($\dot a <0$, $\ddot a <0$) is 
effective to this purpose 
\cite{9}, as expected from the fact that accelerated expansion can 
be transformed into accelerated contraction through 
an appropriate field redefinition, like the one connecting the String and 
the Einstein frame. Indeed, physical effects such as the dilution of 
inhomogeneities should be independent from the choice of the frame. 
So, inflation can be generally identified, in a frame-independent way, 
as a period of accelerated evolution of the scale factor (sign $\dot a=$ 
sign $\ddot a$). 

There are, however, two  classes of accelerated evolution very 
different from a dynamical point of view, and depending on the 
behaviour -- growing or decreasing in time -- of the curvature of the
space-time manifold. A first  important point to be stressed is thus the
fact that the pre-big bang scenario corresponds to a phase of 
accelerated  evolution characterized by growing (or non-decreasing)  
curvature \cite{3,4}, while the phase that we shall call 
``standard inflation " \cite{0,4a} is characterized by  a  curvature  scale
which tends to be constant -- in the limiting  case of a de Sitter metric --  
or slightly decreasing in time. 

Before proceeding further, two remarks are in order. The first is that a
phase of  accelerated evolution and growing curvature, also called
superinflation  (or pole-inflation) \cite{12}, is not a peculiarity of string
cosmology,  but  is possible even in general relativity:  in 
higher-dimensional  backgrounds, for instance, in the context of dynamical
dimensional reduction. The  important difference is that, in string
cosmology, superinflation does not  necessarily requires neither the
shrinking of the internal dimensions, nor some exotic matter source
\cite{11a} or symmetric breaking mechanism \cite{11b}: it can be  simply
driven, even in three spatial dimensions, by the kinetic energy  of the
rolling dilaton field $\phi$ \cite{3,4}, parametrizing the growth of the 
string coupling $g=\exp (\phi/2)$ from zero (the string perturbative 
vacuum), to the  strong coupling regime, $g \sim 1$. 

The second remark is that the words ``pre-big bang" should be referred 
to the complete cosmological evolution from the initial
state approaching the string perturbative vacuum, up to  the beginning of
the hot, radiation-dominated phase of the standard 
scenario. Altough in most of this paper we shall restrict  our discussion to
the low-energy part of the pre-big bang phase, appropriately described in
terms of the low-energy string effective action, the  complete pre-big
bang history necessarily includes a high-curvature ``stringy" phase 
\cite{13}, which may also  be of the inflationary type \cite{17b}, and
which has a curvature  expected to be, on the average, non 
decreasing. 

For a power-law, accelerated, conformally flat background, the
time-behaviour of the curvature scale follows the behaviour of the
absolute value of the  Hubble parameter, $|H| =| \dot a /a|$. In such a
background, on the other hand, the inverse of the  Hubble parameter (i.e. 
the  Hubble horizon $|H|^{-1}$) also controls the (finite) proper
distance between the surface of the event horizon and the wordline of a
comoving observer.  Such a distance is shrinking for pre-big bang inflation
(where  $|H|$ is growing), while it is  non-decreasing in standard inflation. 

As a consequence of the fact that the horizon is shrinking, and the 
curvature is growing, it turns out that the initial state of the phase 
of inflation, in the pre-big  bang scenario, is characterized by a curvature
which is very small in  Planck (or string) units, and by a Hubble horizon very
large   in the same units (for the sake of simplicity we may identify, 
at the end of inflation,   the present value of the string length $L_s$ and of
the Planck lenght  $L_p$; indeed, at tree-level, they are related by
\cite{12a} $L_s=\langle g\rangle L_p= \langle \exp \phi/2 \rangle L_p$,
with a present dilaton expectation  value $\langle g \rangle \sim
0.3-0.03$).  The initial state is also characterized by another very small 
dimensionless number, the initial string coupling $g \ll 1$,
corresponding to a  dilaton approaching the perturbative vacuum
value, $\phi \ra -\infty$. 

By contrast, the standard inflationary scenario is characterized by a 
dilaton already settled  to its present vacuum expectation value; 
the coupling is always strong ($g$ is of order one), and the size  of 
the initial horizon may be of order one in Planck units, if the initial 
curvature approaches the Planck scale. These kinematical and dynamical 
differences between standard and pre-big bang inflation are summarized 
in Table I. 

As evident from the pre-big bang curve of Fig. 2, the smaller is the 
value of the initial curvature, the longer is the duration of the phase 
of pre-big bang inflation. The request that the inflation phase be long 
enough to solve the horizon and flatness problems \cite{7} thus imposes 
bounds on the dimensionless parameters of the initial state, controlled 
by the curvature and by the string coupling. Parameters such as the
typical size of an initial  homogeneous domain, in Planck units, must be by 
far greater than one for inflation to be successful. This  aspect of string
cosmology was  pointed out already in the first  papers on the pre-big
bang scenario \cite{3,4} and, even before, also in the context of
string-driven superinflation \cite{2}.

\begin{table}
\tabcolsep .07cm
\renewcommand{\arraystretch}{2.0}
\begin{center}
\begin{tabular}{|c||c||c|}
\hline
   &  {\bf Standard Inflation}  &  {\bf Pre-big bang Inflation}   \\ \hline

Time evolution     &   {\sl accelerated}  & {\sl accelerated}     \\ \hline

Driving energy   & {\sl inflaton potential}   &  
{\sl dilaton kinetic energy}   \\ \hline

Curvature    &  {\sl constant or decreasing} & {\sl growing}  \\ \hline 

Event horizon    &  {\sl constant or growing} & {\sl shrinking}  \\ \hline 

Initial curvature scale   &  {\sl model-dependent} & {\sl arbitrarily small}  
\\ \hline 

Initial coupling  &  {\sl strong, non-perturbative} & 
{\sl arbitrarily weak, perturbative} \\ \hline 

\end{tabular}
\bigskip
\caption{ Kinematical and dynamical differences 
between standard and pre-big bang inflation.}
\end{center}
\end{table}

The fact that the initial curvature and coupling are 
small, and that the initial state is characterized by very large 
dimensionless numbers, may be interpreted  however as a possible
fine-tuning of the  pre-big bang models \cite{14}, or even as a serious
drawback, preventing  the solution of the flatness and homogeneity
problems, and supporting  the conclusion that ``the current version of the
pre-big bang scenario  cannot replace usual inflation" \cite{15}. 

Consider, for instance,  the horizon/homogeneity problem. At the onset of
pre-big bang  inflation, as usual in an  inflationary context, the 
maximal allowed homogeneity scale is bounded by the size of the horizon. 
The point is that  the initial horizon size $H^{-1}$ is very large, instead of
being of Planckian order. The  basic question thus becomes: is an initial
homogeneity scale of the order of the maximal scale $H^{-1}$  necessarily
unnatural if the initial curvature  is small and,  consequently, the
initial horizon is large in Planck (or string) units?   In other words, which
basic length scale has to be used to measure the  naturalness of  the
initial homogeneous domain, which subsequently  inflates to reproduce the
presently observed Universe? The Planck length  or the radius of the causal
horizon? 

The choice of the Planck length, emphasized in \cite{15}, is certainly 
appropriate when the initial conditions are imposed in a state approaching 
the high-curvature, quantum gravity regime, like  in models of chaotic 
inflation  \cite{16}, for instance. In the pre-big bang scenario, on the 
contrary, the initial conditions are to be imposed when the Universe is 
deeply inside the low-curvature, weak coupling, classical regime. In 
that regime the Universe does not know about the Planck length, and the 
only available classical scale of distance, the horizon, should not be 
discarded ``a priori" as unnatural \cite{21a}. 

This does not mean, of course, 
that the horizon should be always {\sl assumed} as the natural scale of 
homogeneity. This suggests, however,  that the naturalness of
homogeneity over  a large horizon scale should be discussed on the
grounds of some  quantitative and objective criterium, as attempted for
instance in \cite{17}, taking  into account also the effects of quantum
fluctuations \cite{18} that  could destroy the initial, classical homogeneity
(see also \cite{17a} for a discussion of ``generic" initial conditions in a
string cosmology context). 

Another point concerns the flatness problem. In order to explain the 
precise fine-tuning of the present density to the critical one, the 
initial state of pre-big bang inflation must be characterized by large 
dimensionless parameters, thus reintroducing from the beginning the 
large numbers that one would like to explain. 

This may seem to be quite  unsatisfactory, as emphasized in  
\cite{15}.  It should be pointed out, however, that if one accepts the point
of view  that large numbers are always to be avoided at the onset of
inflation,  then should also accept the fact that natural initial 
conditions are only possible in the context of models in which inflation 
starts at the Planck scale, in order to have, for the curvature, an initial  
dimensionless ratio of order one. This rules out, as a satisfactory
explanation of our present  cosmological state, not only the pre-big bang
scenario, but any model in  which inflation starts at scales smaller than
Planckian (unless we imagine a scenario with different stages of inflation,
each of them responsible for solving different problems, and occurring at
different scales: but, again, is this a  natural cosmological configuration?)

Even for a single stage of inflation, occurring very near to the Planck
scale, we are not free from problems, however, as we  are led eventually
to the following question: can we trust the naturalness of inflation 
models in which classical general relativity is  applied to set up initial
conditions at Planckian curvature scales, i.e.  deeply inside the
non-perturbative, quantum gravity regime? (in string cosmology, the
Planckian regime directly affects the exit from the inflationary phase, and
only indirectly set constraints on the initial conditions, through the finite
duration of the low energy phase). 

Assuming that the answer be positive, we are led to a situation that can 
be graphically summarized, in a qualitative way,  as in Fig. 3. 

\begin{itemize}

\item{} 
Case (a) represents a standard inflationary model of the Universe in 
which inflation starts at the Planck scale. The time arrow points from 
bottom to top, and the shaded area at the time $t_0$ represents a 
spatial section of our present homogeneous Universe, of size fixed by 
the present Hubble radius $H_0^{-1}$. As we go back in time, according 
to the solutions of the standard cosmological model, the horizon shrinks 
linearly in cosmic time, while the proper size of the present 
homogeneous region, controlled by the scale factor $a(t)$, shrinks 
slowly. When we reach the Planck scale, at the time $t_f$, the causal 
horizon is smaller than the homogeneous region, roughly by the factor 
$10^{-30}$. 

To solve this problem, the phase of standard evolution is
preceeded by a phase  of exponential de Sitter inflation, long enough
in time from $t_i$ to $t_f$, during which  the curvature and the horizon
stay frozen at the Planck scale, and our  present portion of the Universe
may ``re-enter" inside the causal horizon. 

It should be stressed that, in a realistic inflationary model,  
the horizon has to be slightly increasing from $t_i$ to $t_f$, because 
the scale corresponding to our present Hubble radius has to cross the 
horizon, during inflation, at a curvature scale $H_1$ smaller than
Planckian. We must require, in particular, that   $H_1/M_p~ \laq~ 
10^{-5}$ in order to avoid too much amplification of gravitational 
perturbations, that would contradict the present degree of 
homogeneity observed by COBE \cite{19} at large angular scales. 
However, for the sake of graphical  simplicity,  we shall ignore this
complication that is not  essential for our present discussion. 

\item{}
Case (b) represents a string cosmology model of the Universe, in which 
the inflationary pre-big bang phase, from $t_i$ to $t_f$, is represented 
in terms of the contracting metric of the Einstein frame \cite{9}, in 
order to emphasize (graphically) 
the underlying duality and time-reversal 
symmetry of the scenario (there is no need, of course, that in the 
Einstein frame the pre-big bang scale factor exactly coincide with the 
time-reversal of the post-big bang solution). 

The main difference from 
case (a) is that in the pre-big bang epoch the curvature is growing, and 
the event horizon shrinks linearly in cosmic time, from $t_i$ to $t_f$, 
instead of being constant. Since the scale factor shrinks at a slower 
rate, however, it is still possible for the initial homogeneous domain 
to be ``pushed out"  of the horizon, and for the Universe to emerge at 
the Planck scale, at the time $t_f$, in the same configuration as in 
case (a). The subsequent evolution from $t_f$ to $t_0$ is the same as in 
the standard scenario. 

\item{}
Case (c), finally, represents a string cosmology model in which the period
of pre-big  bang inflation corresponds in part to a phase of growing
curvature,  growing dilaton and shrinking horizon (from $t_i$ to $t_s$), and
in part  to a phase in which the curvature, the horizon, and eventually  the
dilaton, are frozen at the  Planck scale (from $t_s$ to $t_f$). The initial
horizon is still large  in Planck units, but it is no longer 
reflection-symmetric to $t_0$,  depending on the duration of the high
curvature phase from $t_s$ to  $t_f$. 

\end{itemize}

\begin{figure}
\centerline{\epsfxsize=9.0cm
\epsffile{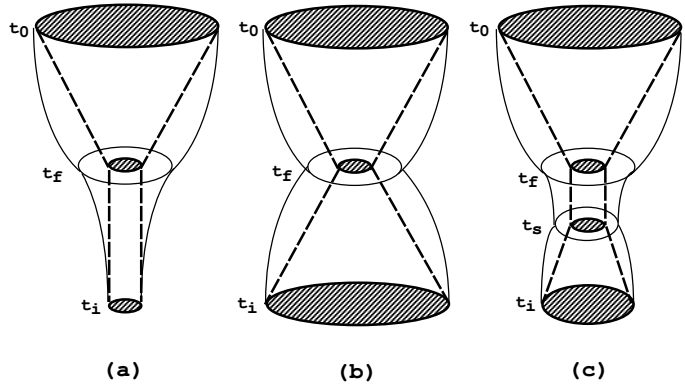}}
\caption{\sl Qualitative evolution of the horizon scale and of the 
proper size of a homogeneous region for (a) standard de Sitter 
inflation with constant Hubble horizon, (b) pre-big bang inflation 
(in the Einstein frame) with 
shrinking horizon, and (c) pre-big bang including a 
phase of high curvature inflation at the string scale. 
The time  
direction coincides with the vertical axis. The three horizontal
spatial sections $t_0, t_f$ and $t_i$ 
corresponds, from top to bottom, to the present
time, to the end, and to the beginning of inflation. The shaded 
area represents the horizon,  and the dashed lines its time 
evolution. The full curves represent the time evolution of the proper size 
of the homogeneous region, controlled by the scale factor.}
\label{fig3}
\end{figure}

According to \cite{15}, the model (a) represents an acceptable solution 
to the horizon and flatness problem because inflation starts at the 
Planck scale, and all the dimensionless parameters characterizing the
initial  configuration are of order one. When the phase at constant
curvature of the model (c) extends in time like in case (a) the two
models  pratically coincide for what concerns the naturalness of the
initial  conditions, as in both cases our Universe emerges at the Planck 
scale from a single domain of Planckian size, and we loose any 
observational tracks of what happened before. 

The aim of the pre-big bang scenario, on the other hand, is to attempt a 
description of the possible cosmological 
evolution {\sl before} the Planck epoch. The main difference between 
case (a) and (c) is that, if the duration of the phase of inflation at 
constant curvature is shorter than the minimal duration required
for a solution of all the standard kinematic problems,  what happened
before the  Planck scale may then become visible. In other words,   there
are  phenomenological consequences that can be ascribed to the phase of
pre-Planckian evolution, and that can be tested (at least in principle)  even
today (see Section \ref{IV}). The model of case (b), in particular, is the
limiting  case in which the duration of the high curvature phase shrinks to
a  point, and the Universe emerges at the Planck curvature scale with a
homogeneous  domain large enough to fill our present Hubble radius
through the subsequent standard evolution. 

The above discussion 
refers to the curvature in Planck units, but the same arguments can also 
be applied to the initial value of the string coupling. If the 
coupling  is of  
order one at the beginning of inflation then it is natural, in the sense 
of \cite{15}, if it is much smaller than one, then the inflationary growth of
the coupling may have observable consequences. 

To conclude this discussion it seems difficult, in our 
opinion, to discard a model of pre-Planckian evolution, like those  
illustrated in case (b), (c), only on the grounds of the large parameters 
characterizing its initial conditions. Such an  argument could be applied in
the impossibility of  observational tests, namely in the absence of any
phenomenological  evidence about the cosmological evolution before the
Planck era. But, as  pointed out before, the pre-Planckian epoch becomes
invisible only in  the limiting case (a), in which the effective models
reduces to standard  inflation  starting at the Planck scale, with no
unnatural initial  conditions. 

If, on the contrary, the high curvature phase
is shorter  than in case (a), then the model requires an initial horizon larger
than  Planckian, and a small initial coupling -- which are possibly unnatural
according to  standard criteria \cite{15}. In that case, however, such initial 
conditions  are in principle accessible to present observations, so why do
not try to test  the scenario observationally, and try to analyze the
naturalness in  terms of a Bayesan approach, as attempted in \cite{17}? In
that case, the  computation of ``a posteriori" probabilities suggests that
the observation  of a large initial horizon and a small initial coupling may
become ``a  posteriori" natural \cite{17}, because of the duality symmetry
intrinsic  to the pre-big bang scenario. 

Finally, even if the initial state should require a certain degree of
fine-tuning, this does not necessarily implies  that the pre-big bang
cosmology described by string models is to be discarded (after all, the
description of our late-time Universe given by the standard cosmological
model is rather satisfactory, in spite of the fine-tuning required in such a
model if the initial state is extrapolated back in time until the Planck
epoch).  

Usually, the need for fine-tuning in the initial conditions means
that the model is incomplete, and that a more general dynamical
mechanism is required,  to explain the particular initial conditions.
Thus, it might well be that the pre-big bang picture provided by  string
cosmology  does not represents the whole story of our Universe, and that
only an  earlier evolution can explain why, at a certain instant of time, the
Universe is lead to a state so similar to the string perturbative vacuum.

\section {Quantum cosmology differences}
\label{III}

In the standard inflationary scenario the phase of exponential, 
de Sitter-like expansion at constant curvature cannot be infinitely 
extended towards the past, for a well known reason  of geodesic
completeness. A complete manifold requires an earlier contracting phase:
in that case, however, it seems impossible, in models dominated by the
inflaton potential, to stop the collapse and to bounce towards the
expanding phase \cite{5}. One has thus the problem of explaining how the
Universe could   emerge at the Planck scale in the initial state appropriate
to exponential expansion. 

At curvature scales of order one in Planck units we are in the full 
quantum gravity regime, and the use of the quantum cosmology approach  
seems to be appropriate. In this approach the Universe is represented by 
a wave function satisfying the Wheeler-De Witt (WDW) equation \cite{20}, 
and evolving in the so-called superspace, whose points represent 
all possible spatial geometric  configurations. For pratical applications,
however, the evolution of the WDW wave  function is usually studied in a
``minisuperspace" context, where only a finite numbers of coordinates
is chosen to parametrize the  different geometrical configurations. 

With an appropriate choice of the boundary conditions it is then  possible
to  obtain cosmological solutions of the WDW equation describing the
``birth  of the Universe" as an effect of quantum tunnelling \cite{21,22}. 
In that case, if the geometric state of the Universe is characterized by a 
cosmological constant $\La$ (due, for instance, to the vacuum 
energy-density induced by a scalar field potential), the tunnelling
probability  is found 
to be proportional to $\exp \left(-\La^{-1}\right)$, where $\La$ 
is measured in Planck units. The Universe tends thus to emerge in a 
state of big vacuum energy, just appropriate to the onset of inflation. 
The quantum cosmology approach seems thus to provide a natural
mechanism  to explain the formation of ``baby universes",  emerging at
the  Planck scale, and ready to inflate according to the standard 
inflationary scenario  \cite{23}. 

The minisuperspace approach to quantum cosmology is known  to be
affected  by various problems of technical nature: the probabilistic 
interpretation, the unambiguous determination of an appropriate time 
parameter, the semiclassical limit, the ordering of quantum operators, 
and so on. The most unsatisfactory aspect of this approach, in our 
opinion, is however the fact that the boundary conditions for the 
tunnelling process are to be chosen ``ad-hoc". They are by no means 
compelling, and it is possible indeed to impose different boundary 
conditions, for instance according to the ``no boundary" criterium 
\cite{24}, leading to a completely different result for the probability 
of creation of universes -- results that are not always appropriate
to inflationary initial  conditions. 

The source of this problem is the fact that, in a quantum description of 
the birth of the Universe, the final cosmological state (i.e. the 
Universe that we want to obtain) is well known, while the initial 
cosmological state (before the quantum transition) is completely 
unknown, at least in the context of the standard inflationary scenario. 
Indeed, the cosmological tunnelling is usually referred as a proces of 
tunnelling ``from nothing" \cite{21}, just to stress the ignorance about 
the initial vacuum state. The classical theory of the standard 
cosmological scenario cannot help, because the initial 
state, in that context,   
is the big bang singularity, i.e. just what the quantum approach 
would like to avoid. 

In a string cosmology context, the  quantum approach based on 
minisuperspace can be implemented in a straightforward way, with the 
only difference that the differential WDW equation represents the 
Hamiltonian constraint following not from the Einstein action, but from the
low-energy string effective  action \cite{25,26,27}. As a  consequence, the
``minimal" minisuperspace is at least two-dimensional,  because the action
always contains the dilaton, besides the metric. The  formal problems
related to the minisuperspace approach remains, with the possible
exception of the operator-ordering problem, as the quantum  ordering  is
unambiguosly fixed by the  global, pseudo-orthogonal $O(d,d)$ symmetry of
the low-energy string  effective action \cite{25,27}. Another  possible
exception is the identification of the time-like  coordinate in
minisuperspace \cite{33a}. 

There is, however, a radical difference for what concerns boundary 
conditions. In the context of the pre-big bang scenario the initial,  
asymptotic state of the Universe is unambiguosly prescribed -- the 
string perturbative vacuum -- and cannot be chosen ``ad-hoc". Such 
initial state is perfectly appropriate to a low-energy normalization of 
the WDW wave function, and the transition probability of string 
cosmology only depends on the dynamics, i.e. on the effective potential 
appearing in the WDW equation. 

It is now interesting to observe that if we compute, in the context of 
the pre-big bang scenario, the transition probability from the 
perturbative vacuum to a final, post-big bang configuration 
characterized by a non-vanishing cosmological constant, we obtain 
\cite{26} a probability distribution $P(\La)$ very similar to the one of the
conventional quantum cosmology, computed with tunneling boundary 
conditions. The reason is that, by imposing the perturbative vacuum as 
the boundary condition to the WDW equation, the WDW solutions contain
only  outgoing waves at the singular spacetime boundary, just like in
the case of  tunnelling boundary conditions \cite{21}. In this sense, we can
say that  the ``ad-hoc" prescription of tunnelling boundary conditions
simulates,  in a phenomenological way, the birth of the Universe from the
string  perturbative vacuum. This suggests that, instead of ``tunnelling
from  nothing", we should speak of ``tunnelling from the string
perturbative  vacuum" or, even better, of quantum instability and ``decay"
of the  perturbative vacuum \cite{28}. 

A further, important difference should be mentioned. The 
transition from the pre- to the post-big bang phase induced by the
cosmological  constant (or, more generally, by an appropriate dilaton
potential), is  represented, in the minisuperspace of string cosmology,  not 
like a  tunnelling effet, but like a quantum reflection of the WDW wave
function,  over an effective potential barrier. The correct description that
we obtain  in  string cosmology  for the birth of the Universe, therefore, is
that of  a  ``quantum scattering" effect \cite{25,26,33a}. 

The various differences between quantum inflationary cosmology and 
quantum string cosmology are summarized in Table II. Besides the formal 
aspects (such as tunnelling versus reflection), the basic difference is 
that in the standard inflationary scenario the Universe, because of 
quantum cosmology effects, is expected {\em to enter} in the 
inflationary regime, while in the pre-big bang scenario the Universe is 
expected {\em to exit} from the inflationary regime (or at least from 
the phase of growing curvature).  So, in standard inflation, quantum 
effects at the Planck scale are expected to be responsible for 
inflationary initial conditions. In string cosmology, on the contrary, 
initial conditions are to be imposed in the opposite, low energy quantum 
regime, where quantum effects are negligible. 

\begin{table}
\tabcolsep .07cm
\renewcommand{\arraystretch}{2.0}
\begin{center}
\begin{tabular}{|c||c||c|}
\hline
   &  {\bf Quantum cosmology}  &  {\bf Quantum string cosmology}  
\\ \hline  

Formal  approach    &   {\sl  WDW, minisuperspace}  & 
{\sl  WDW, minisuperspace}     \\ \hline

 Quantum ordering & {\sl arbitrary }   &  
{\sl fixed by duality}   \\ \hline

Boundary conditions    &  {\sl arbitrary } & {\sl string perturbative vacuum} 
\\ \hline 

Outgoing waves  &  {\sl tunnelling from nothing } & {\sl
reflection from pre-big bang}  \\ \hline 

Quantum transition    &  {\sl beginning of inflation} & {\sl
exit from inflation}   \\ \hline 

\end{tabular}
\bigskip
\caption{ Quantum cosmology differences 
between standard and pre-big bang inflation.}
\end{center}
\end{table}

It seems appropriate, at this point, to comment on the fact that the 
initial state of pre-big bang inflation seems to be characterized by a 
large entropy $S$, if one assumes, as  in \cite{15}, that the de 
Sitter relation between entropy and horizon area remains valid also when 
the horizon is not constant in time (or, in other words, if one assumes a
saturation of the bound provided by the holographic principle \cite{34a},
applied however to the Hubble horizon \cite{34b}). If $S$ is large in Planck
units,  the probability that such a configuration be obtained through a
process  of quantum tunnelling, $\exp (-S)$, is exponentially small, as 
emphasized in \cite{15}.  However, as stressed above, in string cosmology
quantum effects such as  tunnelling or reflection are expected to be
effective  {\em at the end} of inflation, and not  {\em at the
beginning},  i.e. {\em not}  to explain the origin of the initial state. A large
entropy of  the initial state, in the weakly-coupled, highly-classical regime,
can  only correspond to a large probability of such configuration, which is 
proportional to $\exp (+S)$, like for every classical and macroscopic 
configurations (not arising from quantum tunnelling).

Let us stress, finally, another important difference between
conventional  quantum cosmology and quantum string cosmology. In string 
cosmology quantum geometrical effects cannot be fully accounted for, 
as fas as we limit ourself to a WDW equation obtained from the 
low-energy string effective action. Indeed, when approaching the Planck
scale, the  string theory action acquires (even at small coupling, i.e. at 
tree-level in the quantum loop expansion) higher curvature correction 
\cite{29}, weighed by the inverse string tension $\alpha'$. They are to  be
included into the Hamiltonian constraint, and lead  in general to a 
higher-derivative WDW equation. This problem has been discussed in 
\cite{30}, and it has been shown in \cite{31} that when the higher 
curvature corrections appear in the form of an Eulero density, then the
WDW  approach can only be applied to a dimensionally reduced version of
the  theory. 

The quantum cosmology results reported in this section  refer, in 
this sense, only to a model of ``low-energy"  quantum string cosmology
\cite{28}. In the full quantum gravity regime,  in order to include all the
higher-derivative contributions, the correct  WDW equation should follow
not from the effective action, but possibly  from a conformal, sigma-model
action \cite{43a}, which automatically takes into  account all orders in
$\alpha'$.  We note, however, that duality transformations in toroidal
moduli space have recently suggested \cite{30a} that the Lorenztian
structure of the low-energy minisuperspace may have an exact meaning
also in an  $M$-theory context, even if the exact $M$-theory equations are
expected to be in general different. 

In fact, when the curvature is large in string units ($\ap H^2 >1$), and also
the string coupling is large ($g^2=e^\phi >1$), we necessarily enter the
$M$-theory regime where new quantum effects are possible, such as a
copious production of higher-dimensional $D$-branes \cite{M1}. If the
curvature is small enough, the strong coupling regime of string
cosmology is then expected to be described by  $11$-dimensional
supergravity theory, and the dilaton to be interpreted as the radius (i.e.,
the modulus field) of the $11$-th dimension \cite{M2}. In this context
string cosmology becomes $U$-duality covariant \cite{10a, M3,M4},
and the presence in the action of Ramond-Ramond fields may be helpful to
evade the problem of the curvature singularity \cite{M5,M6,M7}. The
singularity, in addition, could also disappear as a result of the embedding
of the low-energy solutions of string theory into a higher-dimensional
($d=11$) manifold \cite{M4,M8}. 

These results seem to suggests that an appropriate quantum description
of the birth of the Universe will be probably achieved only within a full
$M$-theory approach to the strong coupling regime, in which the pre-big
bang acceleration is damped, the curvature is regularized, and the Universe
bounces back to the phase of standard evolution.

\section {Phenomenological differences}
\label{IV}

One of the most important (and probably also most spectacular) 
phenomenological predictions of inflation is the 
parametric amplification of metric (and of other different types of) 
perturbations \cite{32}, and the corresponding generation of primordial 
inhomogeneity spectra, directly from the quantum fluctuations of the 
background fields in their vacuum state (see \cite{33} for a review).  
Such fluctuations, when decomposed in Fourier modes, satisfy a 
canonical Schrodinger-like equation, whose effective potential is 
determined by the  so-called`` pump field", which depends in its
turn on the  background geometry. 

It is then evident that different backgrounds lead to different pump 
fields, to a different evolution of perturbations, and thus to different 
spectra. In string cosmology, in particular, there are two main 
properties of the background that can affect the final form of the 
perturbation spectra. They are:

\begin{itemize}

\item{} 
(A) the growth of the curvature scale;

\item{}
(B) the scalar-tensor (i.e. gravi-dilaton) nature of the background. 

\end{itemize}
Property (A) has two important consequences. The first, that we will call 
(A1), is that the pre-big bang scenario leads to metric perturbation 
spectra growing with frequency \cite{34} (instead of being 
flat, or decreasing,  
like in standard inflation), because the spectral distribution of metric 
perturbations tends to follow the behaviour of the curvature scale 
at the time of the first horizon crossing. The second, that we will call 
(A2), is that the growth of the curvature can also force the comoving 
amplitude of perturbations to grow (instead of being frozen) outside the 
horizon. This effect, implicitly contained in the  earlier, pioneer studies
\cite{41a}, was first explicitly  pointed out in \cite{35}, and only later
independently  re-discovered in a string cosmology context  \cite{36}. As
a further  consequence of property (A) we should mention, finally, the fact
that  perturbations are amplified in a final ``squeezed vacuum" state, and
not  in a ``squeezed thermal vacuum" \cite{36a}. 

Let us first discuss the second effect (A2). This effect, on one hand, is
interesting because it may lead to an amplification of
perturbations more efficient than in the standard  inflationary scenario. On
the other hand, however, it is dangerous,  because the perturbation
amplitude could grow too much, during the  pre-big bang phase, so as to
prevent the application of the standard  linearized approach, which
neglects effects of back-reaction \cite{33}. 

Such an ``anomalous" growth of perturbations cannot be eliminated 
by a change of frame, because the associated 
physical (i.e. observable) energy density spectrum is obviously 
frame-independent. However, the breakdown of the linear approximation 
{\em is}, in general, {\em gauge-dependent}. For the particular case of 
scalar metric perturbations, in three isotropic dimensions, the linear 
approximation breaks down in the standard longitudinal gauge, but is 
restored in a more appropriate ``off-diagonal" gauge \cite{37}, also 
called ``uniform-curvature" gauge \cite{38}. Moreover, as a consequence 
of a particolar form of duality invariance that appears explicitly in 
the Hamiltonian approach to perturbation theory \cite{39}, the final 
energy-density spectrum can  always be correctly estimated  by
neglecting  the growing mode, provided one includes in the full Hamiltonian
both the  contribution of the amplitude and of its conjugate momentum. 

There are backgrounds, however, in which the growth of perturbations 
remains too strong even after the elimination of all unphysical gauge 
effects, and we have to limit ourselves to a restricted portion of 
parameter space for the linear approximation to be valid. Even in this 
case, however, the effect (A1) has two interesting consequences. 

The first is that a  growing spectrum leads to the 
formation of relic backgrounds whose amplitude is higher at higher
frequency,  where in general the backgrounds are also more easily
detectable. A typical example  is the formation of a relic background of
cosmic gravitons which, in the frequency range of present resonant-mass
and interferometric detectors  ($10^2-10^3$ Hz), could be up to eight
orders of magnitude  stronger than expected in the context of standard
inflation \cite{34,40}. Thus in principle detectable, in a not-so-far future, 
by the  (planned) advanced version of the interferometric gravitational
antennas, or by spherical resonant detectors. 

The second consequence is that the  normalization of the peak of the
spectrum,  at high frequency, is  automatically controlled by the string
scale \cite{41}. The peak  amplitude may be high enough to support a
picture in which all the  radiation, that becomes dominant at the beginning
of the standard era, is  produced through a process of parametric
amplification, directly from  the quantum fluctuations of the vacuum
during the pre-big bang epoch \cite{41,42,50a}.  
Indeed, for  the background fields that interact more strongly 
than gravitationally,  the amplified fluctuations are expected  to
thermalize, and their energy-density is expected to grow in  time with
respect to the dilaton kinetic energy that was driving the  background
during the phase of   pre-big bang inflation. This possibility is absent in 
the standard inflationary scenario, where the spectrum of perturbations 
is decreasing, and the normalization of the spectrum is determined at low 
frequency by the observation of the large scale CMB anisotropy 
\cite{19}:  the resulting energy-density of the fluctuations, in that case, is
by far  too low to dominate, eventually, the post-inflationary background. 

Up to now we have reported some phenomenological consequences of the 
property (A). 
For what concerns the property (B), i.e. the gravi-dilaton nature of the 
background, we will quote here a peculiar string cosmology effect, the 
amplification of the quantum fluctuations of the electromagnetic field  
due to their direct coupling to the dilaton, according to the effective 
Lagrangian density $ \sqrt{-g} e^{-\phi} F_{\mu\nu}F^{\mu\nu}$. 

In general relativity the dilaton is absent, the Lagrangian is 
invariant under a conformal rescaling of the metric, and the coupling of 
the electromagnetic field to a conformally flat metric, typical of inflation, 
can  always be eliminated  (unless the coupling is non-minimal and/or
violates $U(1)$ gauge invariance \cite{54a}). As a consequence, the 
inflationary evolution of the metric background is unable to amplify the
electromagnetic fluctuations. In string cosmology, on the contrary, such 
fluctuations are amplified by the accelerated growth of the dilaton   (acting
as the pump field) during the pre-big bang phase. If the  high-curvature
string phase is long enough, it is then possible to produce in  this way the
``seeds", required for instance by the galactic dynamo,  for the generation
of cosmic magnetic fields on a large scale \cite{43}. String  cosmology thus
provides a possible solution to a longstanding  astrophysical ``puzzle", i.e.
the generation of the primordial seed  fields,  through a mechanism  which
is uneffective in the standard inflationary scenario. 

This certainly represents an advantage with respect to the standard 
scenario. The different amplification of perturbations, however, is also 
asssociated to possible drawbacks. In particular, the fact that the 
metric perturbation spectrum, in string cosmology, grows with a very
steep  slope, and it is rigidly normalized at high frequency, makes  
problematic the matching to the anisotropy observed at the present 
horizon scale \cite{19}. The generation of the observed CMB anisotropy, 
with the right spectrum, is instead one of the most celebrated results 
of standard inflation \cite{0,4a,33}. 

A possible solution to this problem, in string cosmology, comes from the 
amplification of the fluctuations of the Kalb-Ramond axion, which is one
of  the fundamental fields appearing in the string effective action, already 
at low energy. Indeed, unlike metric perturbations,  the axion 
perturbations can be amplified with a rather flat spectrum \cite{51a} and,
through the (integrated) Sachs-Wolfe effect \cite{44}, they can  induce the
temperature anisotropy observed in the CMB radiation on a large scale,
both  in case of massless \cite{45} and massive \cite{46} axion fluctuations. 

It is important to observe, in that case, that the slope 
of the spectrum is no longer arbitrary, but rigidly determined by the
COBE  normalization (imposed at low frequency), 
and by the string normalization 
(imposed at the opposite, high-frequency end of the spectrum). The 
resulting slope turns out to be slightly increasing \cite{45,46}, but still in 
agreement with the observational limits at the horizon scale \cite{47}. At
higher frequency scales, however, important   differences from standard
inflation may appear in the peak structure of the spectrum \cite{55a}.
The possible axionic origin of the fluctuations of the CMB temperature is
thus expected to be confirmed (or disproved) in a very near future, by the
planned satellite observations. 

\begin{table}
\tabcolsep .07cm
\renewcommand{\arraystretch}{2.0}
\begin{center}
\begin{tabular}{|c||c||c|}
\hline
   & {\bf Standard Inflation}   &   {\bf Pre-big bang Inflation} \\ \hline  

Pump field   &   {\sl  metric}  & 
{\sl  metric and dilaton}     \\ \hline

Spectrum (vs frequency) & {\sl flat or decreasing }   &  
{\sl increasing}   \\ \hline

Amplitude outside horizon   &  {\sl frozen } & {\sl increasing}  \\ \hline 

Normalization  &  {\sl low frequency (COBE) } & {\sl
high frequency (string scale) }  \\ \hline 

Electromagnetic fields   &  {\sl unaffected } & {\sl
amplified by the dilaton}   \\ \hline 

CMB anisotropy & {\sl metric fluctuations} & 
{\sl axion fluctations} \\ \hline

Dilaton fluctuations & {\sl absent} & 
{\sl dilaton productions} \\ \hline

\end{tabular}
\bigskip
\caption{Amplification of vacuum fluctuations in the inflationary and 
pre-big bang scenario.}
\end{center}
\end{table}

The main differences between standard  and string cosmology inflation,
for what concerns  cosmological perturbations, are  summarized in
Table III. The last entry of the Table refers to the amplification of  the
fluctuations of the dilaton background \cite{48}, another peculiar  effect
of string  cosmology, because the dilaton  is absent in the standard
scenario. This effect, analogous to the  amplification of tensor metric
perturbations, can be interpreted (in a  second quantization approach) as a
process of dilaton production, which  leads to the formation of a relic
background of cosmic dilatons. The background is subject to various
constraints, depending on the slope  of the spectrum and on the mass of
the dilaton, but if dilatons are  light enough  \cite{49} (namely $m ~\laq
~10$ KeV), they are not yet decayed and could  represent today a
significant fraction of the dark matter, that seems  required to match
various astrophysical observations. 

Detecting such a background, through the   gravity-like 
interactions of the dilatons at low-energy scales, is however a challenge 
that seems beyond the possibilities of present technology  \cite{57a},
unless the dilaton couples universally to macroscopic bodies, represents
a significant fraction of dark matter, and its mass lies within the
sensitivity band of gravitational detectors \cite{64a}.

\section {Conclusion}
\label{V}

The pre-big bang scenario provides a model, suggested and 
supported by string theory, of  the possible cosmological evolution before
our Universe emerged at the  Planck scale. 

If the subsequent post-Planckian evolution follows the standard
inflationary scenario,   then  initial conditions can be imposed at the Planck
scale, and  are probably natural in the sense of  \cite{15}, but any track of
what happened before disappears from our observational range. If, on the
contrary, inflation at the Planck scale is not  too long, and  inside our
present  Hubble radius there are comoving length scales that crossed the
horizon  during the low-curvature pre-big bang phase, then pre-Plackian
initial  conditions are in principle accessible to present observations, and
their  naturaleness can be discussed in terms of a Bayesan analysis
\cite{17},   based  on ``a posteriori" probabilities. Quantum cosmology
methods can  also be applied, taking into account however that quantum
effects are  possibly  important at the end, and not
at the beginning,  of inflation. 

We believe  that 
the possibility of looking back in the past before the Planck 
era is the most distinctive aspect of string cosmology,  with respect to
the  standard inflationary cosmology. Concerning the possible tracks of 
the pre-Planckian  Universe, we have emphasized ,  in particular, three
effects, referring to  observations to  be performed $1)$ in a
not-so-far future, $2)$ in a near  future, and $3)$ to observations already
(in part) performed. These  effects are, respectively: $1)$ the production of
a cosmic gravity wave  background, $2)$ the axion-induced anisotropy of
the CMB radiation,  and $3)$ the production of seeds for the cosmic
magnetic fields. 

These effects are not necessarily compatible among them (at least in the 
``minimal" version of the pre-big bang models), and it seems thus possible
to test, and  eventually exclude (or confirm) the pre-big bang scenario on
the grounds  of its phenomenological consequences. Even if, as recently
stressed  \cite{50}, the idea itself of inflation ``cannot be falsified", the
particular models can (and must) be tested, and the pre-big bang scenario,
after all, can  be regarded as a particular, unconventional model of
primordial inflation. 

It seems thus appropriate to stress, in conclusion, that the pre-big bang
scenario is not alternative to the idea of inflation, but only alternative to
a more conventional realization of inflation which, for hystorical reason,
is still deeply anchored to the standard big bang picture, where  the
initial state must necessarily represent a very small, curve and dense
Universe.  The effort of this paper aims at stimulating the reader's 
meditation on the fact that this standard picture is a possibility, not a
necessity, and that quite different initial conditions are possible, and not
necessarily unlikely.

\bigskip
\ack

It is a pleasure to thank Gabriele Veneziano, for the long and fruitful
collaboration on many aspects of the string
cosmology scenario discussed in this paper.

\end{document}